\documentclass[aps,prb,showpacs,eqsecnum, floatfix,superscriptaddress]{revtex4}
\usepackage{epsfig}
\usepackage{graphicx}
\usepackage{dcolumn}
\usepackage{bm}
\usepackage{amssymb}
\usepackage{color}

\newif\ifequation
\equationtrue

\newcommand{\eq}[1]{Eq.~(\ref{#1})}
\newcommand{\fig}[1]{Fig.~{\ref{#1}}}
\newcommand{\tab}[1]{Table~{\ref{#1}}}
\newcommand{\be}{\begin{equation}}
\newcommand{\ee}{\end{equation}}
\newcommand{\bea}{\begin{eqnarray}}
\newcommand{\eea}{\end{eqnarray}}
\newcommand{\pd}{\partial}
\newcommand{\al}{\alpha}
\newcommand{\ba}{\beta}
\newcommand{\la}{\lambda}

\newcommand{\ga}{\gamma}

\begin{document}

\title{Charge confinement and Klein tunneling from doping graphene}

\author{C. Popovici}
\affiliation{Departamento de F\'{\i}sica, Universidade de Coimbra,
3004-516 Coimbra, Portugal}
\affiliation{Departamento de F\'{\i}sica, Instituto Tecnol\'ogico
de Aeron\'autica, 12.228-900, S\~ao Jos\'e dos Campos, SP, Brazil}
\affiliation{Institut f\"ur Theoretische Physik, Universit\"at
Giessen, 35392 Giessen, Germany}
\author{O. Oliveira}
\affiliation{Departamento de F\'{\i}sica, Universidade de Coimbra,
3004-516 Coimbra, Portugal} \affiliation{Departamento de
F\'{\i}sica, Instituto Tecnol\'ogico de Aeron\'autica, 12.228-900,
S\~ao Jos\'e dos Campos, SP, Brazil}
\author{W. de Paula}\affiliation{Departamento de F\'{\i}sica, Instituto
Tecnol\'ogico de Aeron\'autica, 12.228-900, S\~ao Jos\'e dos Campos,
SP, Brazil}
\author{T. Frederico}
\affiliation{Departamento de F\'{\i}sica, Instituto Tecnol\'ogico de
Aeron\'autica, 12.228-900, S\~ao Jos\'e dos Campos, SP, Brazil}


\begin{abstract}
In the present work, we investigate how structural defects in graphene
can change its transport properties. In particular, we show that
breaking of the sublattice symmetry in a graphene monolayer overcomes
the Klein effect, leading to confined states of massless Dirac
fermions.  Experimentally, this corresponds to chemical bonding of
foreign atoms to carbon atoms, which attach themselves to preferential
positions on one of the two sublattices.  In addition, we consider the
scattering off a tensor barrier, which describes the rotation of the
honeycomb cells of a given region around an axis perpendicular to the
graphene layer. We demonstrate that in this case the intervalley
mixing between the Dirac points emerges, and that Klein tunneling
occurs.
\end{abstract}


\maketitle

\section{Introduction and motivation}

Graphene, a monatomic layer of carbon atoms on a honeycomb lattice,
has been synthesized for the first time in 2004
\cite{Novoselov2004,Novoselov2005}. Its remarkable properties, such as
unconventional quantum Hall effect \cite{Gusynin2005,Zhang2005}, Klein
tunneling \cite{Katsnelson2006, Cheianov2006} or charge confinement
\cite{Rozhkov2011}, are mainly a consequence of the fact that at low
energies, the charge carriers in graphene are described by the
relativistic Dirac equation instead of the more familiar Schr\"odinger
equation \cite{Wallace1947,Berger2006}. The single particle dispersion
relation is linear in the momentum, $ E_{k}=\pm v_F |\vec k|$, where
$v_F$ is the Fermi velocity and $\vec k = (k_x, k_y)$ is the fermion
momentum, measured relative to the inequivalent corners of the
Brillouin zone $K$ and $K^{\prime}$, known as Dirac points. The peculiar properties of
graphene triggered immediately a lot of interest, due to its possible
applications in nanoscale devices, but also because graphene is able
to connect different branches of physics. Meanwhile, we have learned
how to produce multilayers of graphene, have started to understand the
effects due to the deviations from a perfect carbon honeycomb
structure and, most important from the point of view of technological
applications, are learning on how to grow samples with structural
defects. Defects can be used to tailor graphene based devices --- see,
for example, the discussions below. However, they can also change the
transport properties of the charge carriers and be at the origin of
the performance deterioration of these devices.  Comprehensive
overviews of graphene can be found in 
Refs.~\onlinecite{CastroNeto:2009zz,Peres2010,Aberggel2010,
Kotov2010,Sarma2011}, and the structural defects are reviewed in
\cite{Banhart2011}. The work reported in \cite{Lahiri2010} gives an
example of an experimental realization of one dimensional (1D) defects
in pure graphene.

Typically, graphene based devices require the ability to confine and
control the charge flow. Confining Dirac fermions is yet a challenging
task, due to the so-called Klein effect \cite{Klein1929,Su1993,
Calogeracos:1999yp}, where a relativistic electron is able to
penetrate a barrier higher than the electron's energy completely
unreflected --- in contrast to the conventional tunneling where the
transmission probability drops off exponentially as the barrier gets
higher. Studies of various types of barriers that lead to Klein
tunneling for electrons in graphene can be found in the literature,
and experimental observations of this effect have been reported
\cite{Huard2007,Gorbachev2008,Beenakker2008,Young2009,Stander2009,
Chaves2010}. Despite the difficulties related to overcoming the Klein
effect, charge confinement in graphene has already been accomplished
experimentally --- see for example Refs.~\onlinecite{Geim2007,
CastroNeto:2009zz,Rozhkov2011} for overviews. Theoretically,
confinement was demonstrated by cutting or bending graphene sheets
\cite{Guinea2010,Pereira2010}, exploiting the transversal degrees of
freedom of the electrons in an electrostatic potential
\cite{Silvestrov2007}, applying magnetic fields \cite{deMartino2007,
Giavaras2009}, deforming the graphene membranes \cite{Fogler2008} or
by spatial modulation of the Dirac gap \cite{Giavaras2010,
Giavaras2011}. From structural defects, confinement has been achieved
in graphene-graphane systems, see for example \cite{Singh2010}.
However, so far it has not been explored a way to produce charge
confinement in association with chemical bonding of foreign atoms to
carbon atoms, although techniques to implement the chemical bonding of
some adsorbents, such as fluorine, hydrogen or oxygen, already exist
\cite{Yan2009,Cheng2010,Xiang2010}.

In addition to the possible technological applications, the 1D defects
in pristine graphene are interesting because they can be modeled as
potential barriers, associated with different fermionic operators in
the Dirac equation. Within the conceptual framework of effective
quantum field theories, the authors of \cite{Oliveira:2010hq}
generalize the results of \cite{Hou:2006qc,Jackiw2007} --- where
vortex formation in graphene is described through a chiral gauge
theory --- and introduce a scalar and a gauge field that account for
the dynamics of the self-interaction of the carbon background and the
mean self-interaction of the Dirac fermions. This model is well suited
to describe various disorder phenomena such as topological defects,
doping defects or distortions of the lattice honeycomb.  Within this
framework, carbon nanotubes and graphene have been studied and their
quantum properties have been reproduced \cite{Cordeiro2009,
Chaves2011}.

In the present study, inspired by the work of \cite{Oliveira:2010hq,
Hou:2006qc,Jackiw2007}, we explore how charge confinement and Klein
tunneling can be induced by certain types of defects, and examine how
defects --- modeled as 1D potential barriers --- can be mapped into
fermionic operators. As described below, charge localization can be
achieved via a barrier which breaks the sublattice symmetry. In
practice, the breaking of the sublattice symmetry can be realized by
binding covalently foreign atoms with particular carbons, whereas for
the theoretical analysis, we have to identify and explore the
corresponding fermionic operators. The Klein effect is investigated by
considering the scattering of an electron off a tensor barrier. The
associated fermionic operator generates a rotation around the $z$-axis
(perpendicular on the graphene plane) and it couples both the two
sublattices $A$ and $B$, and the two valleys $K$ and
$K^\prime$. Experimentally this can be implemented by topologically
distorting the graphene layer via a rotation of the honeycomb cells
around the $z$-axis in a particular region, relative to the remaining
graphene sheet. In order to include the effects of the $K
\leftrightarrow K^\prime$ mixing, we work in the four-component, as
opposed to the standard two-component formalism.  We demonstrate that
for this type of barrier a peculiar effect arises, i.e. the splitting
of the electron wave function into two components inside the barrier,
and we analyze its implications for the solution of the tunneling
problem.

The paper is organized as follows. In Sec.~II we briefly review the
relevant features of the gauge model proposed in
Refs.~\onlinecite{Jackiw2007,Oliveira:2010hq}, concentrating on the
operators that account for the effects of the carbon background and
the carbon-fermion interaction. In this setting, we construct
different types of barriers associated with defects in graphene, and
discuss the implementation of the corresponding operators in the Dirac
equation.  Sec.~III is dedicated to charge confinement. The
transmission probability of an electron scattered off a barrier with
sublattice symmetry breaking is investigated, and the conditions that
enable the clustering of charge are identified.  In Sec.~IV, we focus
our attention on the Klein tunneling, and consider the scattering off
a barrier that describes a spatial distortion of the graphene
sheet. As before, we discuss the angular behavior of the transmission
coefficient, underlining the effects related to the mixing of the
Dirac points. In Sec.~V a short summary and the conclusions will be
presented.

\section{The theoretical setup}

Let us begin by sketching the fermionic dynamics in graphene-based
materials.  Tight binding models provide a first approximation to
describe the electronic properties of graphene, as they only take into
account the fermionic degrees of freedom, whereas the carbon
interactions are summarized via the hopping parameters which control
the electron dynamics.  Since the dynamical degrees of freedom in
graphene must include both the fermions and carbon background, in
\cite{Oliveira:2010hq} a relativistic-like gauge model for graphene
and nanotubes is suggested, which takes into account both the
electron-hole and the carbon dynamics.  In this framework, electrons
are described by a four component Dirac-type spinor, while the carbon
degrees of freedom are associated with a scalar field $\varphi$ and a
gauge field $A^\mu$.  Since the full description is somewhat
long-winded, here we briefly present the main features of the model
and direct the reader to \cite{Oliveira:2010hq} for a full account. In
this gauge model, the electron dynamics is described by the Dirac
equation
\be
\left\{i\gamma^\mu D_\mu - P(\varphi)
- P_5(\varphi) \gamma_5 \right\} \psi = 0,
\label{eq:dirac1}\ee
where the polynomials $P(\varphi)$ and $P_5 (\varphi)$ describe the
interaction between fermions and the carbon crystal structure.
Explicitly, the vacuum expectation value of the scalar field $\varphi$
vanishes for pure graphene, while for doped graphene $\langle \varphi
\rangle = \varphi_0 \ne 0$, where $\varphi_0$ is a minimum of the
scalar potential $V( \varphi^\dagger \varphi)$.  If one is able to
fabricate graphene with islands where it is doped, one can simulated
its electronic properties by taking
\be
P(\varphi)+P_5(\varphi)\ga_5=\left\{\begin{array}{lll}
0 & & \mbox{for undoped regions,} \\ & & \\
(g_2 + i \, h_2 \, \gamma_5) \, \varphi^2_0 & & 
\mbox{for doped regions} \end{array} \right.
\label{eq:SPS}\ee
in the Dirac equation, where $g_2$, $h_2$ are the corresponding
coupling constants.  In this way, one reduces the problem to the
investigation of the solutions of the Dirac equation in a square
potential barrier, which includes the contribution \eq{eq:SPS} and can
in principle account for any combination of Dirac $\gamma$ matrices.
Furthermore, given that the potential barrier is associated with the
doping of graphene, experiment can help to calibrate the numerical
values of $g_2 \varphi^2_0$ and $h_2 \varphi^2_0$.  The covariant
derivative is $D_\mu = \partial_\mu + i g A_\mu$, and the
four-component Dirac spinor is given by
\ifequation\be 
\psi =\left( \begin{array}{c} \psi^b_+\\ \psi^a_ +
\\ \psi^a_- \\ \psi^b_- \end{array} \right), 
\ee \fi 
where the upper index refers to the sublattices $A$ or $B$ and the
lower index to the Dirac points $K$ (plus sign) and $K^\prime$ (minus
sign).  For the Dirac $\gamma$ matrices we will use the representation
\ifequation\be
\gamma^0 = \left( \begin{array}{cc} 0 & I \\ I  & 0 \end{array}
\right) , \quad\gamma^j = \left( \begin{array}{cc} 0 & \sigma^j \\
- \sigma^j  & 0 \end{array} \right),\quad
\gamma_5 = \left( \begin{array}{cc} -I & 0 \\ 0 &  I \end{array}
\right),
\ee\fi
where $\sigma^j$ stands for the $j$ Pauli matrix.  Since we are
concerned with two dimensional fermions, the index $\mu$ takes the
values $\mu=0,1,2$. The Dirac matrices satisfy
$\{\ga^{\mu},\ga^{\nu}\}=2g^{\mu\nu}$.

\begin{table}[t]
\centering\begin{tabular}{l@{\hspace{1cm}}l}
\hline&  Operators \\\hline
(S,S) & $I$, $\gamma_0$, $\gamma_x \gamma_5$ , $\Sigma_x$ \\
(S,A) & $\gamma_5$, $\gamma_x$, $\sigma_{0x}$ \\
(A,S) & $\gamma_y$, $\gamma_z$, $\sigma_{0y}$, $\sigma_{0z}$ \\
(A,A) & $\gamma_y \gamma_5$, $\gamma_z \gamma_5$, $\Sigma_y$,
$\Sigma_z$ \\\hline
\end{tabular}
\caption{The first column refers to the operator symmetry under
interchange of sublattice and intervalley indices, respectively. The
table assumes that the graphene sheet is on the $x-y$ plane. The
$\Sigma$ matrices are defined as $\sigma^{ij} = - i \, \epsilon^{ijk}
\Sigma_k$, where $\epsilon^{ijk}$ is the three dimensional
Levi-Civita, $\epsilon^{123} = 1$ and a sum over the index $k$ is
implicit.}\label{tab:booktabs}
\end{table}

In order to simulate other dynamical effects, which can be associated
with the carbon structure or with the carbon--electron interaction, or
to include the effects due to impurities/defects, new operators should
be added to the Dirac equation (\ref{eq:dirac1}). These operators can
be, in principle, any of the remaining matrices $\gamma^\mu$,
$\gamma^\mu\gamma_5$, $\sigma^{\mu\nu}$ (in this case the values of
$\mu$ extend to the full four-dimensional Minkowski space).  The
symmetry properties of all the 16 operators under interchange of
sublattices and of the Dirac points, respectively, are listed in
\tab{tab:booktabs}. One can proceed and provide an interpretation for
each of the 16 operators of the Dirac algebra in terms of graphene
defects and corresponding barriers. In the following, we will not
provide the detailed form of each contribution; instead, we will
concentrate on the operators that have a simple physical
interpretation, and represent the experimental situations considered
in this work. If one adds a new term to the Dirac equation, let us say
$- \Gamma \psi$, it contributes to the Hamiltonian of the system as
$\overline \psi \, \Gamma \psi$, which can be rewritten in terms of
the spinor components $\psi^a_+$, $\dots$ For the particular choice of
$\Gamma = \gamma^z$ (assuming that $x-y$ is the graphene plane) it
follows that $\overline \psi \, \gamma^z \psi = [
(\psi_{+}^{a})^{\dagger}\psi_{+}^{a} + (\psi_{-}^{a})^{\dagger}
\psi_{-}^{a} ]-[ (\psi_{+}^{b})^{\dagger}\psi_{+}^{b}+
(\psi_{-}^{b})^{\dagger}\psi_{-}^{b} ]$. This term distinguishes the
two carbon sublattices, it is antisymmetric under interchange of the
sublattice indices, and favors the occupation of sublattice $B$
relative to sublattice $A$. Of course, if the coupling constant
associated with this operator is negative, the occupation of
sublattice $A$ is preferred. As discussed previously, the breaking of
the sublattice symmetry can be realized by binding covalently foreign
atoms with particular carbons in graphene and, in this way, a
$\gamma^z$ operator can be simulated. It turns out that this operator
(and its connection with 1D potential barriers) overcomes the Klein
effect and leads to charge confinement. A second operator of interest
for this work is the tensor interaction $\ga^1\ga^2=i
\Sigma_z,\label{eq:tensor}$ where $\Sigma_z$ is the third component of
the spin operator (perpendicular on the plane of the graphene). It
describes the rotation of the graphene around the $z$ axis and can be
related to a topological defect of the lattice in graphene (see for
example \cite{Gonzalez2001} for a study of the low energy properties
of graphene due to distortions of graphene sheets).  Importantly, the
aforementioned operators change the fermion dispersion relation, such
that the electrons acquire an effective mass, with an energy gap twice
the effective mass. For an explicit derivation of the fermion gap, and
its correlation with the underlying scalar field, we refer the reader
to Ref.~\onlinecite{Oliveira:2010hq}.

Since the coupling with the scalar field $\varphi$ devise the
electronic properties of the doped regions of graphene, in the
following we will set the gauge field $A_{\mu}=0$.  After adding the
relevant operators to the Dirac equation (\ref{eq:dirac1}), we obtain:
\ifequation\be
\left\{ i \gamma^\mu \pd_\mu -D_S -i  D_P  \ga^5- D_0
\ga^0- D_V  \ga^z -i D_T \ga^1\ga^2\right\} \psi = 0,
\label{eq:dirac2}\ee\fi
where we have introduced the notations $D_S= P(\varphi)$ and $D_P= -i
P_{5} (\varphi)$ for the scalar and pseudoscalar interactions,
respectively.  $D_{T(V)}$ denotes the strength of the tensor (vector)
interaction, and the term proportional to $\ga_0$ can be viewed as a
chemical potential. Starting with the above equation, the task is now
to investigate the transmission of the electrons through a 1D barrier,
and establish under which conditions the charge confinement and Klein
tunneling take place.

\section{Charge confinement}

In this section we restrict the analysis to a set of operators that
include the scalar, pseudoscalar and vector barriers in the Dirac
equation. Setting $D_T=0$, our starting Dirac equation takes the form
\ifequation\be
\left\{ i \gamma^\mu \pd_\mu -D_S -i \, D_P \, \ga^5- D_0 \,
\ga^0- D_V \, \ga^z\right\} \psi = 0.
\label{eq:dirac3}\ee\fi

\begin{figure}
\centering\includegraphics[width=2in]{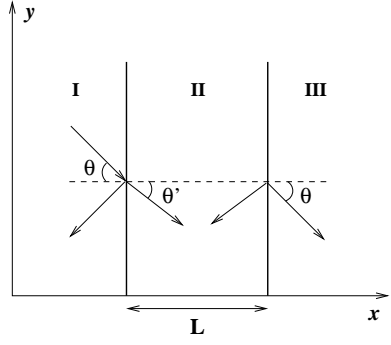}
\caption{Sketch of a one-dimensional defect in an infinite slab of
graphene. $\theta$ and $\theta^{\prime}$ are the angles used in the
scattering problem in regions $I$ (pure graphene), $II$ (graphene with
defect), $III$ (pure graphene), and $L$ is the width of the barrier.
\label{fig:defeito}}
\end{figure}
The scattering over an electrostatic barrier (within the two-component
formalism) was already investigated in
Ref.~\onlinecite{Katsnelson2006} for single layer and double layer
graphene, and the Klein paradox has been demonstrated. When one takes
into account in the Dirac equation the pseudoscalar interaction
$\bar\psi \, \gamma_5\psi$, the sublattice symmetry breaking operator
$\bar\psi \, \gamma^z\psi$ and the scalar term $\bar\psi\,\psi$, the
scattering pattern computed in Ref.~\onlinecite{Katsnelson2006}
changes dramatically. In particular, the tunneling associated with the
Klein effect can be avoided, leading to charge confinement.

Following \cite{Katsnelson2006}, we will consider the simple geometry
of an 1D defect in graphene, where the doping occurs only for the
region for $0 < x < L$.  The ``experimental'' setup is illustrated in
\fig{fig:defeito}.  The wave function describing an incident electron
with energy $E$ and momentum $\vec{p} = E ( \cos\theta \, \hat{e}_x +
\sin\theta \, \hat{e}_y)$, in the region $I$ ($x <0$), is given by
\ifequation \be
\psi_I (x) =  e^{i \vec{p} \cdot \vec{x}}
\left( \begin{array}{c} A_K \\  - A_Ke^{i\theta}\\
B_{K^\prime}e^{-i\theta}\\B_{K^\prime} \end{array}\right)
+   e^{i \vec{p}^{\,\prime} \cdot \vec{x}}
\left( \begin{array}{c} R_K \\ R_K e^{-i\theta}\\
 - R_{K^\prime} e^{i\theta}\\ R_{K^\prime} \end{array}\right) ,
 \label{Scalar_Sol_I}\ee\fi
where the momentum $\vec{p}^{\,\prime} = E ( -\cos\theta \, \hat{e}_x
+ \sin\theta \, \hat{e}_y)$ corresponds to the solution associated
with the reflected wave in region $I$.  For $0 < x < L$ one has
\begin{eqnarray}
\psi_{II} (x) = e^{i \vec{q} \cdot \vec{x}}\left\{ a_K \left(
\begin{array}{c} 1\\ 0 \\ \mathcal{D}_+
\\ \mathcal{D}_q \, e^{i \theta^\prime}\end{array}\right)+b_{K^\prime}
\left( \begin{array}{c} 0 \\1\\ \mathcal{D}_q \, e^{-i \theta^\prime}
\\ \mathcal{D}_- \end{array}\right) \right\}
+  e^{i \vec{q}^{\,\prime} \cdot \vec{x}} \left\{ c_K \left(
\begin{array}{c} 1 \\ 0 \\ \mathcal{D}_+
\\ -\mathcal{D}_q e^{-i \theta^\prime}\end{array}\right)+ d_{K^\prime}
\left( \begin{array}{c} 0 \\ 1\\ - \mathcal{D}_q e^{i \theta^\prime}
\\ \mathcal{D}_- \end{array}\right) \right\},
\label{Scalar_Sol_II}\end{eqnarray}
with
\ifequation \be
\mathcal{D}_\pm = \frac{E-D_0 \pm D_V}{D_S + i D_P} \quad
\mbox{ and }\quad\mathcal{D}_q = \frac{|\vec q|}{D_S + i D_P}.
\ee\fi
As before, $\vec{q} = |\vec q| ( \cos \theta^\prime \, \hat{e}_x +
\sin\theta^\prime \, \hat{e}_y)$ and $\vec{q}^{\,\prime} = |\vec q| 
(- \cos \theta^\prime \, \hat{e}_x + \sin\theta^\prime \, \hat{e}_y)$.
The Dirac equation has a nontrivial solution in region $II$ if the
corresponding determinant vanishes, which leads to the condition
\ifequation \be
\vec q^{\,\, 2} = (E - D_0)^2-D_V^2 - D_S^2 - D_P^2.
\label{eq:condq}\ee\fi
As shall shortly become clear, it is important to notice that the
component $q_x=\sqrt{\vec q^2-E^2\sin^2\theta}$ can be real or a pure
imaginary number, depending on the parameters entering \eq{eq:condq}
and the angle of incidence. For a real $q_x$ the Klein tunneling is
allowed and the barrier is perfectly transparent for certain angles of
incidence. On the other hand, when $q_x$ is imaginary the barrier
becomes opaque and transmission is highly suppressed.  For $x > L$ the
wave function is given by
\ifequation \be
\psi_{III} (x) = e^{i \vec{p} \cdot \vec{x}}
\left( \begin{array}{c} t_K \\ - t_K e^{i\theta}\\
t_{K^\prime} e^{-i\theta}\\t_{K^\prime} \end{array}\right) .
 \label{Scalar_Sol_III}\ee\fi

\begin{figure*}\begin{center}
\includegraphics[width=1.0\linewidth]{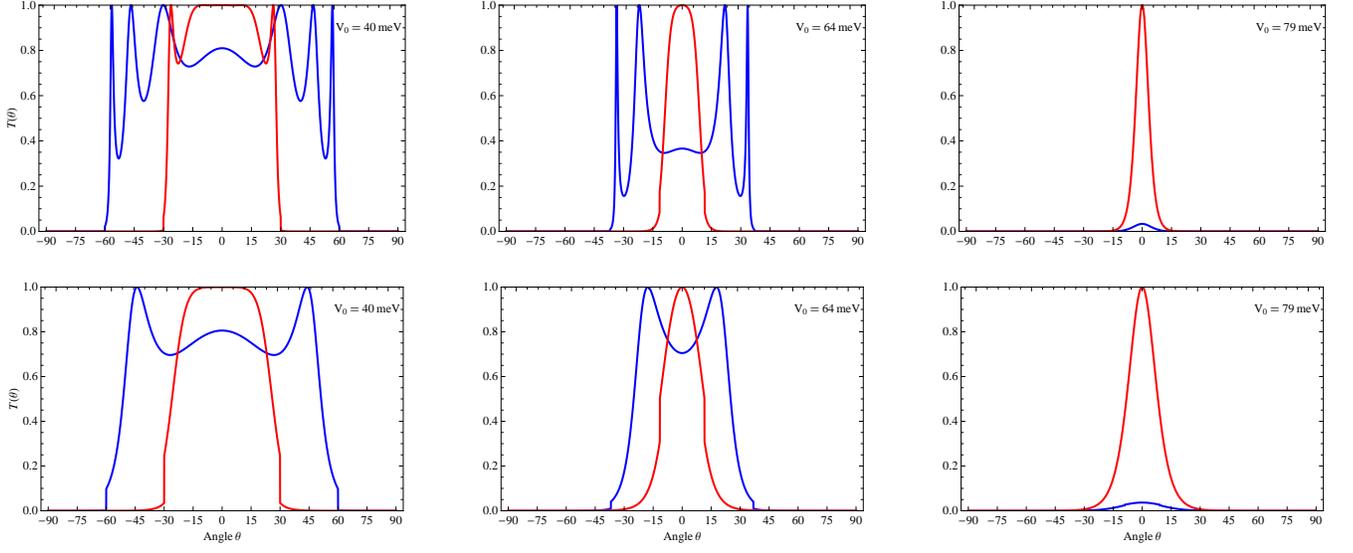}
\caption{(Color online) Angular behavior of the transmission
probability $T(\theta)$ for a $\gamma^z$ barrier (blue line) plotted
against a $\gamma^0$ barrier (red line), for the incoming energy $E =
80$ meV and different heights of the barrier. The widths of the
barriers are $L = 110$ nm (upper panel) and $L = 50$ nm (lower panel).
In the figures, $V_0$ denotes the height of the barriers, i.e.
$V_0=D_0$ for the $\gamma^0$ barrier and $V_0 = D_V$ for the
$\gamma^z$ barrier (in general, in the second case $V_0 = \sqrt{D^2_S
+ D^2_P + D^2_V}$ if one takes into account all barriers associated
with graphene doping; see also text for details). The $\ga^0$ barriers
reproduce the results found in the two-component formalism, as
described in Refs.~\onlinecite{CastroNeto:2009zz, Katsnelson2006}.
\label{fig:BarrierV}}
\end{center}\end{figure*}
For a relativistic system the wave function is continuous everywhere,
therefore $\psi_I(x=0) = \psi_{II}(x=0)$ and $\psi_{II}(x=L) =
\psi_{III}(x=L)$. From the matching conditions and with the definition
of the transmission probability
\ifequation \be
T(\theta)=\frac{|t_K|^2+|t_{K^\prime}|^2}{|A_K|^2+|B_{K^\prime}|^2},
\label{eq:transmcoeff1}\ee\fi
one obtains the following expression:
\ifequation \be
T(\theta)=\frac{\vec q^2\cos^2\theta \cos^2\theta^\prime}
{\vec q^2\cos^2(q_x L)\cos^2\theta\cos^2\theta^\prime
+\sin^2(q_x L)(E-D_0-|\vec q|\sin\theta\sin\theta^\prime)^2},
\label{eq:transmcoeff}\ee\fi
where $\theta^\prime=\arctan (p_y/q_x)$. In the above formula, we
notice that the the Dirac points $K$ and $K^\prime$ do not mix, in the
sense that terms of the type $A_K B_{K^\prime}^{*}$ are not present.
Further, note that the transmission probability corresponding to the
$\ga^0$ operator, with $D_S=D_P=D_V=0$, reproduces the solution
obtained in Ref.~\onlinecite{Katsnelson2006} within the two-component
formalism (stemming from the four-component formalism, when the
couplings between components are discarded). We remind the reader that
in this case the result (\ref{eq:transmcoeff}) reduces to:
\ifequation \be
T_0(\theta)=\frac{\cos^2\theta \cos^2\theta^\prime}
{\cos^2(q_x L)\cos^2\theta\cos^2\theta^\prime
+\sin^2(q_x L)(1-s \sin\theta\sin\theta^\prime)^2},
\label{eq:transmcoeff0}\ee\fi
where $s=\textrm{sgn} (E-D_0)$.

Since the $\ga_0$ interaction is not relevant for our purpose to
demonstrate charge confinement, in the following we will set $D_0=0$.
The potentials $D_S$, $D_P$ and $D_V$ all generate an electronic
massive-type dispersion relation, see \eq{eq:condq}. In this sense,
the doping associated with any of the scalar, pseudoscalar and $\ga^z$
operators change the inertia of the charge carriers in a similar way.
More, given that the $I$, $\ga_5$ and $\ga^z$ contributions in
\eq{eq:condq} are symmetric, it is enough to study one of the three
barriers, say that associated with the $\gamma^z$ interaction $D_V$
(which, as discussed, breaks the sublattice symmetry), while setting
$D_S$ and $D_P$ to zero. Then the condition \eq{eq:condq} simplifies
to \ifequation \be \vec q^2= E^2-D_V^2. \label{eq:condq1} \ee\fi
The apparent singularities in the wave function, \eq{Scalar_Sol_II},
can be discarded simply by multiplying the spinor with a factor
$(D_S+i D_P)$, which changes its length but does not affect the
physical results. Moreover, we have checked that setting $D_S = D_P
=D_V= 0$ directly in the Dirac equation leads to the transmission
coefficient $T_0(\theta)$, \eq{eq:transmcoeff0}, although the
solutions of the Dirac equation inside the barrier look rather
different.

In \fig{fig:BarrierV} the angular behavior of the transmission
probability for a $\gamma^z$ barrier, \eq{eq:transmcoeff}, with $\vec
q^2$ given by \eq{eq:condq1}, (blue line) is plotted against the
$\gamma^0$ barrier, \eq{eq:transmcoeff0} (red line), for various
heights and for two barrier lengths. As shown in the left panels of
\fig{fig:BarrierV}, for a $\gamma^z$ barrier lower than the energy of
the incident electron, there are several directions for which the
barrier is transparent. This is the ``standard'' Klein tunneling ---
already demonstrated for a $\gamma_0$ barrier \cite{CastroNeto:2009zz,
Katsnelson2006} --- which occurs due to the fact that the wave function
inside the barrier, \eq{Scalar_Sol_II}, has an oscillating behavior
(the momentum $|\vec q|$ is real in this case), just like the wave
function outside the barrier. Considering normal incidence
($\theta=\theta^\prime=0$) and keeping $D_V<E$, we notice that the
$\gamma^z$ barrier is not perfectly transparent, as opposed to the
$\gamma^0$ barrier. Instead, transparency occurs only for higher
angles of incidence. On a $\gamma^0$ barrier, for 
$\theta=\theta^\prime=0$, $T_{0}(0)=1$ for any value of $q_x L$,
independent of the height of the barrier, as can be seen from
\fig{fig:BarrierV} and from \eq{eq:transmcoeff0}, whereas in a
$\gamma^z$ barrier, $T(0)$ is no longer unit but depends on the ratio
$E/D_V$. A straightforward calculation gives
\ifequation\be
T(0)= \frac{1}{1 + \frac{D^2_V}{E^2 - D^2_V}\sin^2(q_x L)}\ee\fi
and therefore $T(0) < 1$, for $D_V<E$.  Increasing the barrier height,
the range of angles where the perfect tunneling through a $\gamma^z$
barrier is allowed shrinks until the Klein tunneling is suppressed, as
can be seen in the last plots of \fig{fig:BarrierV}. This behavior can
be understood by looking at \eq{eq:transmcoeff}. Indeed, setting $E =
D_V$, it follows that $\vec q^2 = 0$ and $T(\theta) = 0$ for any angle
of incidence. When the height of the barrier is higher then the
electron energy, the momentum $\vec q^{\,\, 2}$, \eq{eq:condq1}, flips
the sign such that $\sqrt{\vec q^{\,\, 2}}$ and the component $q_x$
become imaginary, and the wave function inside the barrier switches
from oscillating to exponential decay. In turn, this implies that for
barriers higher than the electron energy, the barrier becomes opaque,
thus leading to the charge clustering of the electrons.

\section{Klein tunneling and intervalley mixing}

So far, in the scattering off a barrier of scalar, pseudoscalar and
vector type, the transitions $K \leftrightarrow K^\prime$ did not
occur, in the sense that a term that mixes the coefficients
corresponding to the two different valleys was not present in the
transition amplitude. In the present section, we investigate the
effects on Klein tunneling emerging from the $K \leftrightarrow
K^\prime$ transitions, by introducing a tensor term proportional to
$\gamma^1\gamma^2=i\Sigma_z$ (and discard all other contributions),
where $\Sigma_z$ is the third component of the spin operator in the
four-spinor representation. Recall that such an operator can be
associated with a rotation of the honeycomb carbon lattice. In this
case, the Dirac equation (\ref{eq:dirac2}) takes the form
\be
\left\{ i \gamma^\mu \pd_\mu -i D_T \ga^1\ga^2\right\} \psi = 0.
\label{eq:dirac0}\ee

As before we will consider the geometry of an almost 1D defect in
graphene, such that the electron is scattered off a barrier that
ranges between $0$ and $L$ (see also \fig{fig:defeito}). For an
incident electron moving with energy $E$ and momentum $\vec{p} = E
(\cos\theta \,\hat{e}_x + \sin\theta \, \hat{e}_y)$, the wave
functions in the regions $I$ ($x<0$) and $III$ ($x>L$) are given by
Eqs.~(\ref{Scalar_Sol_I}) and (\ref{Scalar_Sol_III}).
Inside the barrier, where the contribution of the tensor interaction
is non-vanishing, a plane wave nontrivial solution for the Dirac
equation requires \be D_{T}^2=(E\pm |\vec q|)^2.\ee For a given
electron energy $E > 0$ there are two possible momenta, $|\vec
q^{(1,2)}|=\left| E\pm D_{T} \right|$, with the associated spinors
given by, respectively,
\bea
\psi^{(1)}=\left( \begin{array}{c} 1\\-e^{i \theta^{(1)}}
\\-1\\  -e^{i \theta^{(1)}}\end{array} \right),\textrm{~~}
\psi^{(2)}=\left( \begin{array}{c} 1\\-e^{i \theta^{(2)}}
 \\1\\  e^{i \theta^{ (2)}}\end{array} \right).
\label{eq:tensor_spinors}
\eea
This means that, in the scattering through the 1D tensor barrier, the
incoming wave is divided inside the region II, see
Fig.~\ref{fig:defeito}, into two distinct plane waves, propagating
with different momenta and at different angles, $\theta^{(1)}$ and
$\theta^{(2)}$.  Since one wave is exponentially enhanced and the
other, exponentially suppressed, their mixing will produce a pattern
that exhibits perfect tunneling for several angles, regardless of the
height of the barrier. This is in contrast to the $\gamma_{z}$ type of
defect considered in the previous section, where for a barrier higher
than the electron energy the transmission probability goes very
rapidly to zero.  Using \eq{eq:tensor_spinors}, we can write down the
general solution of the Dirac equation for $0<x<L$:
\be \psi_{II} (x) =
e^{i \vec{q}^{\,(1)} \cdot \vec{x}}\al_K \,
\left(\begin{array}{c} 1\\ -e^{i \theta^{(1)}} \\-1\\ 
-e^{i\theta^{(1)}} \end{array}\right) +
e^{i \vec{q}^{\,(2)} \cdot \vec{x}} \ba_{K^\prime} \,
\left( \begin{array}{c} 1\\-e^{i \theta^{(2)}} \\ 1\\
e^{i \theta^{(2)}}\end{array}\right) +
e^{i \vec{q}^{\,\prime (1)} \cdot \vec{x}}\eta_K \,
\left( \begin{array}{c} 1\\ e^{-i \theta^{(1)}} \\-1\\
e^{-i \theta^{(1)}}  \end{array}\right) +
e^{i \vec{q}^{\,\prime (2)} \cdot \vec{x}}\la_{K^\prime} \,
\left( \begin{array}{c} 1\\e^{-i \theta^{(2)}} \\ 1\\
-e^{-i \theta^{(2)}}\end{array}\right),
\label{Tensor_Sol_II}\ee
where $\vec{q}^{\,(1,2)}=|\vec{q}^{\,(1,2)}| e^{i \theta^{(1,2)}}$ and
$\vec{q}^{\,\prime (1,2)}=-|\vec{q}^{\,(1,2)}| e^{-i \theta^{(1,2)}}$.

As in the previous section, we use the fact that a solution of the
Dirac equation is a continuous function everywhere, in order to obtain
the boundary conditions for the wave function $\psi$.  For the tensor
barrier, it is convenient to write the boundary conditions at the
borders of region $II$ in matrix form. At $x=0$ we have
\bea \left(\begin{array}{cccc}
1& 0&1& 0\\-e^{i \theta} &0 &e^{-i \theta}&0\\
0&1 &0&1\\0&e^{i \theta} &0&-e^{-i \theta}
\end{array}\right)\left(\begin{array}{c}
A_K\\B_{K^\prime}\\R_{K}\\ R_{K^\prime}
\end{array}\right)=\left(\begin{array}{cccc}1& 1&1& 1\\
-e^{i \theta^{(1)}} &-e^{i \theta^{(2)}} &e^{-i \theta^{(1)}}
&e^{-i \theta^{(2)}} \\-1&1 &-1&1\\
-e^{i \theta^{(1)}}&e^{i \theta^{(2)}} &e^{-i \theta^{(1)}}
&-e^{-i \theta^{(2)}} \end{array}\right)
\left(\begin{array}{c}
\al_K\\\ba_{K^\prime}\\\eta_{K}\\ \la_{K^\prime}
\end{array}\right).
\label{eq:boundary0} \eea
With the notations
\bea M=\left(\begin{array}{cccc} 1& 0&1& 0\\
-e^{i \theta} &0 &e^{-i \theta}&0\\0&1 &0&1\\
0&e^{i \theta} &0&-e^{-i \theta}
\end{array}\right)\mathrm{~and~~}
N=\left(\begin{array}{cccc} 1& 1&1& 1\\
-e^{i \theta^{(1)}} &-e^{i \theta^{(2)}} &e^{-i \theta^{(1)}}
&e^{-i \theta^{(2)}} \\-1&1 &-1&1\\
-e^{i \theta^{(1)}}&e^{i \theta^{(2)}} &e^{-i \theta^{(1)}}
&-e^{-i \theta^{(2)}} \end{array}\right),
\eea
we can rewrite the above equation as
\be M \left(\begin{array}{c}
A_K\\B_{K^\prime}\\R_{K}\\ R_{K^\prime}
\end{array}\right)= N \left(\begin{array}{c}
\al_K\\\ba_{K^\prime}\\\eta_{K}\\ \la_{K^\prime}
\end{array}\right).
\label{eq:boundary1} \ee
Similarly, the boundary condition in $x=L$ reads:
\be 
e^{i p_x L} M\left(\begin{array}{c}
t_K\\t_{K^\prime}\\r_{K}\\ r_{K^\prime}
\end{array}\right)=N\left(\begin{array}{cccc}
e^{i q_{x}^{(1)}L}&0&0&0\\
0&e^{i q_{x}^{(2)}L}&0&0\\
0&0&e^{-i q_{x}^{(1)}L}&0\\
0&0&0&e^{-i q_{x}^{(2)}L}
\end{array}\right)
\left(\begin{array}{c}
\al_K\\\ba_{K^\prime}\\\eta_{K}\\ \la_{K^\prime}
\end{array}\right).
\label{eq:boundaryD} \ee
Notice the insertion of the coefficients $r_K, r_{K^\prime}$ in the
transmitted wave. Even though this alteration does not change the
final result, there are two reasons that motivate it.  Firstly, with
this modification the matrix that multiplies the outgoing spinor
becomes identical to the matrix $M$ in \eq{eq:boundary1},
corresponding to the incoming wave --- otherwise, if
$r_K=r_{K^\prime}=0$ the matrix $M$ should be modified such that the
last two columns are zero, although formally they can be left
unchanged since the multiplication with zero does not change the final
result. A second (physical) reason is related to the possibility of
investigating a double (or multiple) square potential, where the wave
emerging from the first barrier does have a reflected component, which
is then scattered on a second barrier, and so forth.  With the
notation
\bea \tilde L=\left(\begin{array}{cccc}
e^{i q_{x}^{(1)}L}&0&0&0\\
0&e^{i q_{x}^{(2)}L}&0&0\\
0&0&e^{-i q_{x}^{(1)}L}&0\\
0&0&0&e^{-i q_{x}^{(2)}L}
\end{array}\right), \eea
and putting Eqs. (\ref{eq:boundary1}), (\ref{eq:boundaryD}) together,
we find that the relation between the coefficients corresponding to
the incoming and outgoing waves can be written in matrix form as (here
we set $r_K, r_{K^\prime}$ back to zero and remove the phase factor
$e^{i p_x L}$, which does not contribute to the transition amplitude)
\be \left(\begin{array}{c} t_K\\t_{K^\prime}\\0\\0
\end{array}\right)=M^{-1}N\tilde LN^{-1}M
\left(\begin{array}{c}
A_K\\B_{K^\prime}\\R_{K}\\ R_{K^\prime}
\end{array}\right). \label{eq:matr}\ee
Furthermore, in order to calculate the transmission probability, one
can rewrite the above equation in a form that eliminates the
dependence on the reflection coefficients $R_{K}, R_{K^\prime}$.  With
the notation \be X=M^{-1}N\tilde L N^{-1}M,\label{eq:matrX}\ee one can
split the equation (\ref{eq:matr}) into ``blocks'', and rewrite it in
the following form:
 \bea \left(
\begin{array}{c}t_K\\t_{K^\prime}\end{array} \right)
=X_{11}\left(\begin{array}{c}A_K\\B_{K^\prime}
\end{array}\right)+X_{12}\left(\begin{array}{c}R_K\\R_{K^\prime}
\end{array}\right),\\
\left(\begin{array}{c}0\\0\end{array}\right)
=X_{21}\left(\begin{array}{c}A_K\\B_{K^\prime}
\end{array}\right)+X_{22}\left(\begin{array}{c}R_K\\R_{K^\prime}
\end{array}\right), \eea
where $X_{ij}$ are $2\times 2$ matrices, corresponding to the four
entries of the matrix $X$, \eq{eq:matrX}.  It is straightforward to
show that
\bea \left(\begin{array}{c}t_K\\t_{K^\prime}\end{array}\right)
=\left(X_{11}-X_{12}X_{22}^{-1}X_{21}\right)
\left(\begin{array}{c}A_K\\B_{K^\prime}\end{array}\right),
\label{eq:relcoeff} \eea 
and hence we have an explicit relation between the coefficients that
contribute to the transmission probability. We have used $Mathematica$
to tackle our problem and the calculation shows a matrix $X_{11}$ with
non-vanishing off diagonal terms, i.e. that produce a mixing between
the coefficients $A_{K}$ and $B_{K^\prime}$.  This implies that a
rotation of the honeycomb carbon lattice induces an intervalley
transition ($K \leftrightarrow K^\prime$) in the barrier reflection
and transmission.  The various momenta and angles entering the above
equations are given by
\bea
&&p_x= E \cos\theta,\; p_y=E \sin\theta,\\
&&q_x^{(1,2)}=|\vec q^{(1,2)}|\cos\theta^{(1,2)},\;
q_y^{(1)}=q_y^{(2)}=p_y=|\vec q^{(1)}|\sin\theta^{(1)} 
=|\vec q^{(2)}|\sin\theta^{(2)},\\
&&\theta^{(1,2)}=\arcsin \frac{E \sin\theta}{E\pm D_{T}},
\label{eq:param} \eea
such that the only free parameters are the incidence angle $\theta$,
the width of the barrier $L$ and the strength of the interaction
$D_{T}$. The transmission coefficient is then calculated numerically,
using the definition \eq{eq:transmcoeff1} and the relation
\eq{eq:relcoeff}.

\begin{figure}
\centering\includegraphics[width=1.0\linewidth]{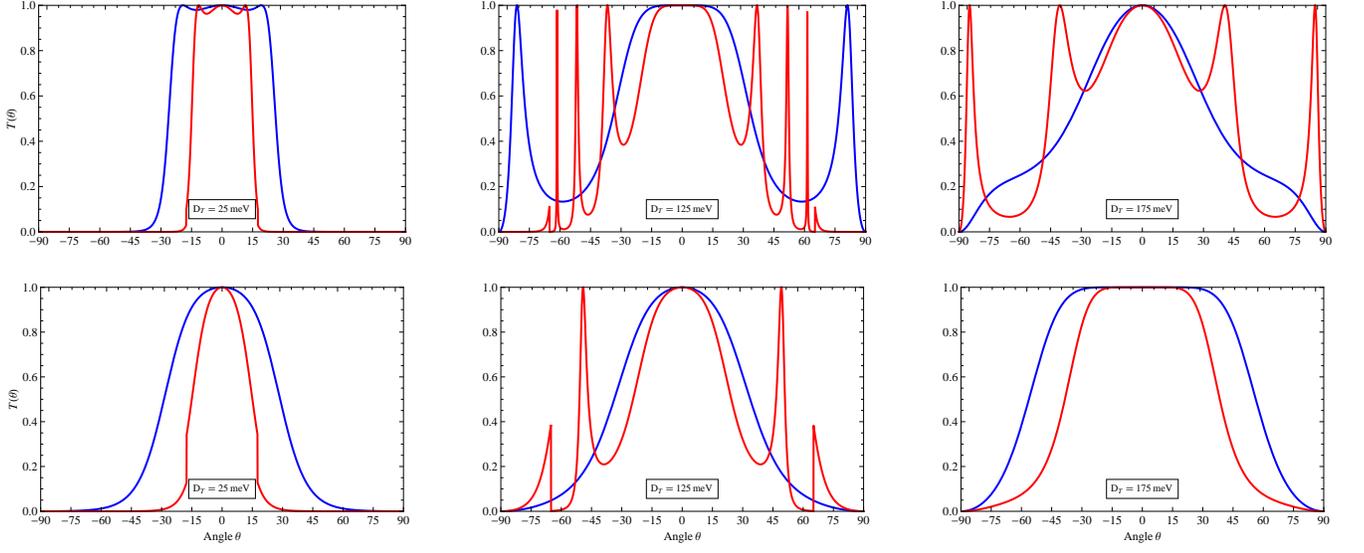}
\caption{(Color online) Angular behavior of the transmission
probability $T(\theta)$ for a tensor barrier
$\gamma^1\gamma^2=i\Sigma_z$ (blue line) plotted against a 
$\gamma^0$ barrier (red line), for the incoming energy $E = 50$ meV
and different heights of the barrier. The widths of the barrier are
$L=110$ nm (upper panel) and $L=50$ nm (lower panel), and $D_T$
denotes the height of the barrier. \label{fig:Barrier3}}
\end{figure}
\begin{figure}
\centering\includegraphics[width=1.0\linewidth]{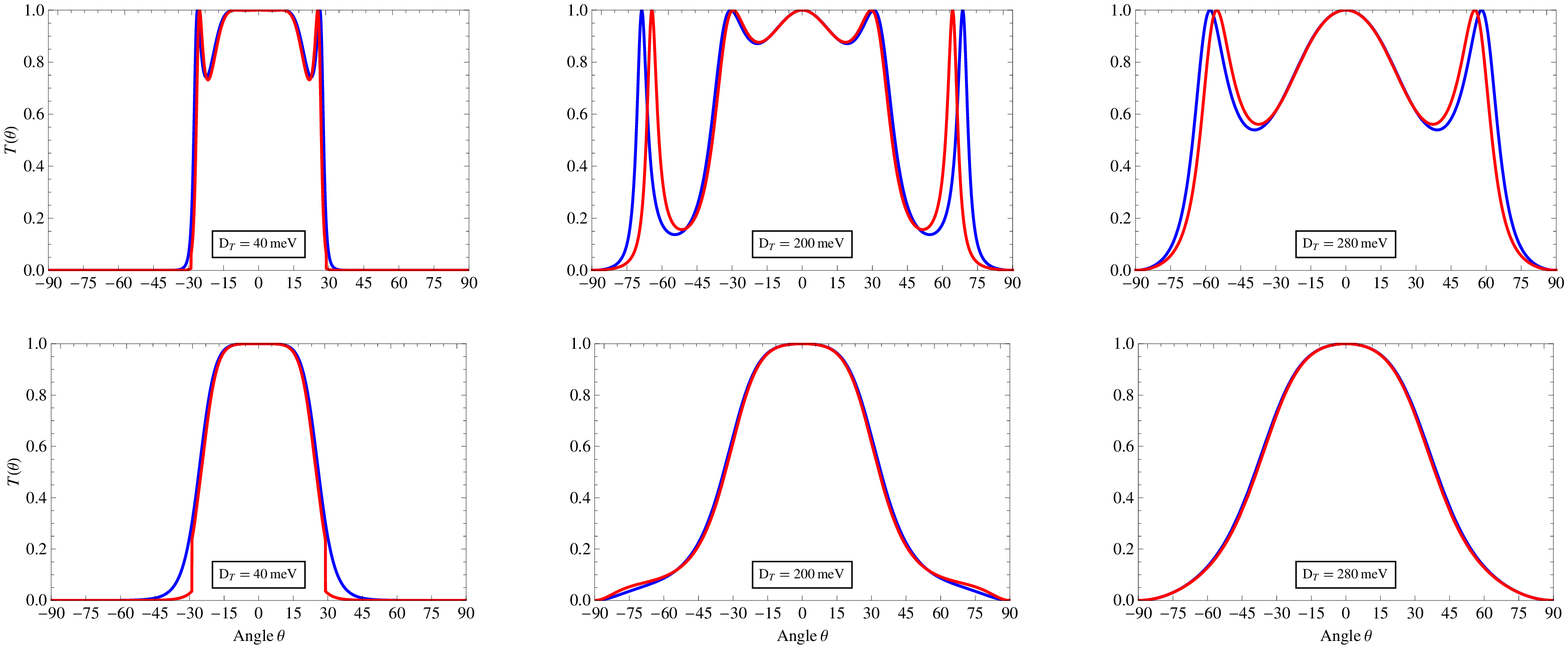}
\caption{(Color online) Same as in \fig{fig:Barrier3}, for an energy
of the incoming electron $E=80$ meV. The widths of the barrier are
$L=110$ nm (upper panel) and $L=50$ nm (lower panel), and $D_T$
denotes the height of the barrier. \label{fig:Barrier2}}
\end{figure}
\begin{figure}
\centering\includegraphics[width=1.0\linewidth]{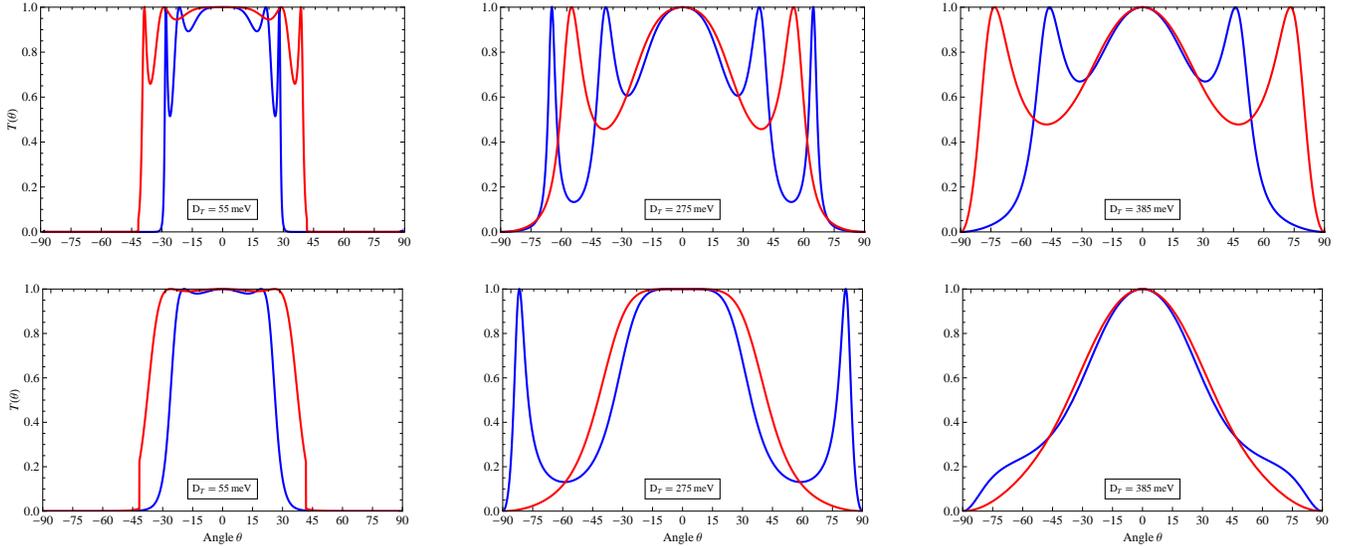}
\caption{(Color online) Same as in \fig{fig:Barrier3}, for an energy
of the incoming electron $E=110$ meV. The widths of the barrier are
$L=110$ nm (upper panel) and $L=50$ nm (lower panel), and $D_T$
denotes the height of the barrier.\label{fig:Barrier1}}
\end{figure}

In order to investigate the Klein tunneling, including the effects of
the $K \leftrightarrow K^{\prime}$ mixing, in
Figs.~\ref{fig:Barrier3}--\ref{fig:Barrier1} we plot the transmission
probability for a tensor barrier (blue line) against a $\ga^0$ barrier
given by \eq{eq:transmcoeff0} (red line).  For a lower value of the
electron energy ($E=50$ meV), we notice that for a relatively low
barrier (half of the electron energy), the tensor barrier has a higher
acceptance than the $\ga^0$ barrier --- this is mostly visible for a
width of $L=110$ nm, where the transmission coefficient is close to
one for $\theta \le 38^o$ for tensor and $\theta \le 17^o$ for $\ga^0$
barrier, as can be seen in \fig{fig:Barrier3}. Increasing the height,
the $\ga^0$ barrier picks up more ``structure'', becoming transparent
for central incidence and at several special angles, whereas the
tensor interaction shows a less rich ``structure''. For even higher
barrier the behavior follows the same trend except that there is a
smaller number of angles for which perfect tunneling is allowed. In
the second set of plots (depicted in \fig{fig:Barrier2}) we increase
the electron energy, while keeping the widths of the barriers
constant.  Choosing the electron energy identical to the value used in
the calculation presented in \cite{CastroNeto:2009zz} ($E=80$ meV), we
find that the transmission probabilities for tensor and $\ga^0$
barriers are almost identical. Finally, inspecting the last set of
plots (\fig{fig:Barrier1}), we find that for even higher electron
energies ($E=110$ meV), the roles of the barriers are reversed: for a
lower height, the tensor barrier has a lower acceptance compared with
the $\ga^0$ barrier, whereas for higher barriers the tensor component
exhibits more angles for which perfect tunneling takes place. Our
results show that the effects of the $K\leftrightarrow K^\prime$
mixing, combined with the splitting of the wave inside the barrier,
are more prominent for higher energies of the incoming electron. We
speculate that this is a result of the fact that the exponential
factors of the wave function inside the barrier, \eq{Tensor_Sol_II},
depend on the absolute value of the electron energy $E$ (and the
barrier height $D_T$), whereas the angles \eq{eq:param} only depend on
the ration $E/D_T$.

\section{Summary and conclusions}

In this paper, we have used the four component spinor formalism to
investigate the transmission coefficients for $\gamma^z$, $\gamma^0$,
$I$, $\gamma_5$ and $\gamma^1 \gamma^2$ one-dimensional barriers in
graphene. We have worked in the framework of the gauge theoretical
model proposed in Refs.~\onlinecite{Jackiw2007,Oliveira:2010hq}, where
the electron-hole and carbon dynamics are simulated by a scalar and
gauge fields.  The different operators in the Dirac equation can be
related to various types of defects in graphene: the $\ga^z$ operator
is related to the breaking of the sublattice symmetry, $\gamma^1
\gamma^2$ describes a topological defect of the graphene sheet,
whereas $\ga^0$ can be associated with a chemical potential, and $I$
and $\gamma_5$ appear in effective gauge theories for graphene
\cite{Jackiw2007,Oliveira:2010hq}.

For the $\gamma^0$ barrier we reproduce the results of the previous
calculations performed within the two component spinor formalism
\cite{Katsnelson2006}. For the $\gamma^z$, $I$ and $\gamma_5$ barriers
we found that the transmittance has a strong dependence on the barrier
height and incidence angle of the electron. In particular, under
certain conditions these interactions generate an opaque barrier which
gives rise to charge confinement. From an experimental point of view,
the most prominent example is the breaking of the sublattice symmetry
associated with the $\ga^z$ operator, which can be implemented via
covalent bonding of foreign atoms to carbon atoms. Hence, according to
our calculation, chemical bonding can be used as an instrument for
controlling charge clustering in graphene. The remaining
one-dimensional barrier, corresponding to the tensor interaction
$\gamma^1 \gamma^2=i\Sigma_3$, exhibits an unusual
feature. Specifically, it appears that the wave function of the
incoming electron is divided into two plane waves inside the barrier,
propagating at different angles and with different momenta.  This
translates into a distinct behavior of the transmission probability as
a function of the incidence angle of the electron: while the Klein
tunneling is observed just like in the case of the $\ga^0$ barrier
\cite{CastroNeto:2009zz, Katsnelson2006}, for particular values of the
energy the scattering patterns of the two barriers can differ
significantly (see also Figs.~\ref{fig:Barrier3}--\ref{fig:Barrier1}).
We hope that understanding the nature of the tunneling states will be
of further use in studies of the transport properties of
graphene-based devices.

\section{Acknowledgments}

The authors acknowledge financial support from the Brazilian agencies
FAPESP (Funda\c c\~ao de Amparo \`a Pesquisa do Estado de S\~ao
Paulo), CNPq (Conselho Nacional de Desenvolvimento Cient\'ifico e
Tecnol\'ogico). OO and CP acknowledge financial support from the
Portuguese agency FCT (Funda\c c\~ao para a Ci\^encia e Tecnologia)
under contract PTDC/\-FIS/100968/2008. CP also acknowledges support
from the Deutsche Forschungsgemeinschaft through contract SFB 634.


\begin{thebibliography}{99}

\bibitem{Novoselov2004}
K.~S.~Novoselov, A.~K.~Geim, S.~V. Morozov, D.~Jiang, Y.~Zhang,
S.~V.~Dubonos, I.~V.~Grigorieva, and A.~A.~Firsov,
Science \textbf{306}, 666 (2004).

\bibitem{Novoselov2005}
K.~S.~Novoselov, A.~K.~Geim, S.~V.~Morozov, D.~Jiang,
M.~I.~Katsnelson, I.~V.~Grigorieva, S.~V.~Dubonos, and A.~A.~Firsov,
Nature (London) , \textbf{438}, 197 (2005).

\bibitem{Gusynin2005}
V.~P.~Gusynin, and S.~G.~Sharapov,
Physical\ Review\ Letters, \textbf{95}, 146801 (2005).

\bibitem{Zhang2005}
Y.~Zhang, Y.-W.~Tan, H.~L.~Stormer, and P.~Kim,
Nature \textbf{438}, 201 (2005).

\bibitem{Cheianov2006}
 V.~V.~Cheianov, and V.~I.~Fal'Ko,
Phys.\ Rev.\ B \textbf{74}, 041403 (2006).

\bibitem{Katsnelson2006}
M.~I.~Katsnelson, K.~S.~Novoselov, and A.~K.~Geim,
Nature\ Physics \textbf{2}, 620 (2006).

\bibitem{Rozhkov2011}
A.~V.~Rozhkov, G.~Giavaras, Y.~P.~Bliokh, V.~Freilikher, and F.~Nori,
Phys.\ Rep.\ \textbf{503}, 77 (2011).

\bibitem{Wallace1947}
P.~R.~Wallace,
Phys.\ Rev.\ \textbf{71}, 622 (1947).

\bibitem{Berger2006}
C.~Berger, Z.~Song, X.~Li, X.~Wu, N.~Brown, C.~Naud, D.~Mayou, T.~Li,
J.~Hass, A.~N.~Marchenkov, E.~H.~Conrad, P.~N.~First, and
W.~A.~de~Heer, Science \textbf{312}, 1191 (2006).

\bibitem{CastroNeto:2009zz}
A.~Castro~Neto, F.~Guinea, N.~Peres, K.~Novoselov, and A.~Geim,
Rev.\ Mod.\ Phys.\ \textbf{81}, 109 (2009).

\bibitem{Peres2010}
N.~M.~R.~Peres,
Rev.\ Mod.\ Phys.\ \textbf{82}, 2673 (2010).

\bibitem{Aberggel2010}
D.~S.~L.~Abergel, V.~Apalkov, J.~Berashevich, K.~Ziegler, and
T.~Chakraborty,
 Adv.\ Phys.\ \textbf{59}, 261 (2010).

\bibitem{Kotov2010}
V.~N.~Kotov, B.~Uchoa, V.~M.~Pereira, A.~H.~Castro~Neto, F.~Guinea,
arXiv:1012.3484 [cond-mat].

\bibitem{Sarma2011}
S.~Das~Sarma, S.~Adam, E.~H.~Hwang, E.~Rossi,
Rev.\ Mod.\ Phys.\ \textbf{83}, 407 (2011).

\bibitem{Banhart2011}
F.~Banhart, J.~Kotakoski, and A.~V.~Krasheninnikov,
ACS\ Nano \textbf{5}, 26 (2011).

\bibitem{Lahiri2010}
J.~Lahiri, Y.~Lin, P.~Bozkurt, I.~I.~Oleynik and M.~Batzill,
Nature\ Nano \textbf{53}, 326 (2010).

\bibitem{Klein1929}
O.~Klein, Z.\ Phys.\ \textbf{53}, 157 (1929).

\bibitem{Su1993}
R.~K.~Su, G.~G.~Siu, and X.~Chou,
Journal\ of\ Physics\ A: Mathematical\ and\
General \textbf{26}, 1001 (1993).

\bibitem{Calogeracos:1999yp}
A.~Calogeracos and N.~Dombey,
Contemp.\ Phys.\  \textbf{40}, 313 (1999).

\bibitem{Beenakker2008}
C.~W.~J.~Beenakker,
Rev.\ Mod.\ Phys.\  \textbf{80}, 1337 (2008).

\bibitem{Chaves2010}
J.~M.~Pereira~Jr, F.~M.~Peeters, A.~Chaves, and G.~A.~Farias,
Semiconductor\ Science\ and\ Technology \textbf{25}, 033002 (2010).

\bibitem{Gorbachev2008}
R.~V.~Gorbachev, A.~S.~Mayorov, A.~K.~Savchenko, D.~W.
Horsell, and, F.~Guinea,
Nano\ Letters \textbf{8}, 1995 (2008).

\bibitem{Young2009}
A.~F.~Young, and P.~Kim,
Nature Physics \textbf{5}, 222 (2009).

\bibitem{Stander2009}
A.~F.~Young, and P.~Kim,
Physical\ Review\ Letters \textbf{102}, 026807 (2009).

\bibitem{Huard2007}
B.~Huard, J.~A.~Sulpizio, N.~Stander, K.~Todd, B.~Yang, and
D.~Goldhaber-Gordon,
Physical\ Review\ Letters \textbf{98}, 236803 (2007).

\bibitem{Geim2007}
A.~K.~Geim and K.~S.~Novoselov,
Nature\ Materials \textbf{6}, 183 (2007).

\bibitem{Guinea2010}
F.~Guinea, M.~I.~Katsnelson, and A.~K.~Geim,
Nature\ Physics \textbf{6}, 30 (2010).

\bibitem{Pereira2010}
V.~M.~Pereira, A.~H.~Castro~Neto, H.~Y.~Liang, and L.~Mahadevan,
Phys.\ Rev.\ Lett.\ \textbf{105}, 156603 (2010).

\bibitem{Silvestrov2007}
P.~G.~Silvestrov and K.~B.~Efetov,
Phys.\ Rev.\ Lett.\ \textbf{98}, 016802 (2007).

\bibitem{deMartino2007}
A.~de~Martino, L.~Dell~Anna, and R.~Egger,
Phys.\ Rev.\ Lett.\ \textbf{98}, 066802 (2007).

\bibitem{Giavaras2009}
G.~Giavaras, P.~A.~Maksym, and M.~Roy,
Journal\ of\ Physics\ Condensed\ Matter \textbf{21},
102201 (2009).

\bibitem{Fogler2008}
M.~M.~Fogler, F.~Guinea, and M.~I.~Katsnelson,
Phys.\ Rev.\ Lett.\ \textbf{101}, 226804 (2008).

\bibitem{Giavaras2010}
G.~Giavaras and F.~Nori,
App.\ Phys.\ Lett.\ \textbf{97}, 243106 (2010).

\bibitem{Giavaras2011}
G.~Giavaras and F.~Nori,
Phys.\ Rev.\ B \textbf{83}, 165427 (2011).

\bibitem{Singh2010}
A.~K.~Singh, E.~S.~Penev, and B.~I.~Yakobson,
ACS\ Nano\ \textbf{4}, 3510  (2010).

\bibitem{Yan2009}
J.-A.~Yan, L.~Xian, and M.~Y.~Chou,
Phys.\ Rev.\ Lett.\ \textbf{103}, 086802 (2009).

\bibitem{Cheng2010}
S.-H.~Cheng, K.~Zou, F.~Okino, H.~R.~Gutierrez, A.~Gupta,
N.~Shen, P.~C.~Eklund, J.~O.~Sofo, and J.~Zhu,
Phys.\ Rev.\ B \textbf{81}, 205435 (2010).

\bibitem{Xiang2010}
H.~J.~Xiang, E.~J.~Kan, S.-H.~Wei, X.~G.~Gong, and M.-H.~Whangbo,
Phys.\ Rev.\ B \textbf{82}, 165425 (2010).

\bibitem{Oliveira:2010hq}
O.~Oliveira, C.~E.~Cordeiro, A.~Delfino, W.~de~Paula, and
T.~Frederico, Phys.\ Rev.\ B \textbf{83}, 155419 (2011).

\bibitem{Jackiw2007}
R.~Jackiw, S.-Y.~Pi,
Phys.\ Rev.\ Lett.\ \textbf{98}, 266402 (2007).

\bibitem{Hou:2006qc}
C.-Y.Hou, C.~Chamon, and C.~Mudry,
Phys.\ Rev.\ Lett.\ \textbf{98}, 186809 (2007).

\bibitem{Cordeiro2009}
C.~Cordeiro, A.~Delfino, and T.~Frederico,
Carbon \textbf{47}, 690 (2009).

\bibitem{Chaves2011}
A.~J.~Chaves, G.~D.~Lima, W.~de~Paula, C.~E.~Cordeiro, A.~Delfino,
T.~Frederico, O.~Oliveira,
Phys.\ Rev.\ B \textbf{83}, 153405 (2011).

\bibitem{Gonzalez2001}
J.~Gonzalez, F.~Guinea, and M.~A.~Vozmediano,
Phys.\ Rev.\ B \textbf{63}, 134421 (2001).

\end{thebibliography}
\end{document}